\documentclass[
 reprint,
 superscriptaddress,
 amsmath,amssymb,
 aps,
 prl,
 floatfix,
]{revtex4-2}

\usepackage{hyperref}
\hypersetup{colorlinks,allcolors=blue}

\usepackage{graphicx}
\usepackage{dcolumn}
\usepackage{bm}
\usepackage{xcolor}

\newcommand{\method}{DPI}

\begin{document}
\raggedbottom

\title{Nuclear Quantum Effects as a Denoising Problem}

\author{Weizhou Wang}
\affiliation{Department of Chemistry, University of Chicago, Chicago, Illinois 60637, USA}
\affiliation{James Franck Institute, University of Chicago, Chicago, Illinois 60637, USA}

\author{Jonathan Weare}
\email{weare@nyu.edu}
\affiliation{Courant Institute of Mathematical Sciences, New York University, New York, New York 10012, USA}

\author{Aaron R. Dinner}
\email{dinner@uchicago.edu}
\affiliation{Department of Chemistry, University of Chicago, Chicago, Illinois 60637, USA}
\affiliation{James Franck Institute, University of Chicago, Chicago, Illinois 60637, USA}

\date{\today}

\begin{abstract}
Nuclear quantum effects are rigorously captured by imaginary-time path integrals, which map the quantum Boltzmann distribution onto a ring polymer of classical replicas. Yet the nuclear masses, the coupling to the environment, and the boundary conditions of the path remain hard-wired in the simulation or the trained model, even though this quantum context enters the path measure only through a quadratic action known in closed form. Here we show that a denoiser trained on classical Boltzmann statistics alone, composed at sampling time with an analytic Gaussian component carrying the entire quantum context, yields the quantum Boltzmann distribution of the nuclei. Such a composition exists and is exact whenever the training noise does not exceed the intrinsic quantum uncertainty of the target ensemble, and it is invariant across all quantum contexts admitted by this bound. We show exact transfer across temperature, isotopic mass, dissipation strength, and the boundary conditions of the path in theory and in numerical experiments, without retraining. The last yields the end-to-end displacement and momentum distributions of a tagged nucleus from open imaginary-time paths. The same invariance extends in principle to the permuted boundary conditions of bosonic exchange, with the identical denoiser. In this view, the noise of generative modeling and the quantum fluctuations of the nuclei are two faces of the same quadratic structure.
\end{abstract}

\maketitle

\section{Introduction}

Zero-point energy and tunneling of light nuclei govern phenomena ranging from the hydrogen-bond network of liquid water to isotope effects and proton transfer in biomolecules. Standard molecular dynamics (MD) treats the nuclei classically, neglecting these effects \cite{markland_nuclear_2018}. The path-integral formulation provides a rigorous framework to map the quantum Boltzmann distribution onto a classical ring polymer of $P$ replicas (beads) of the system connected by harmonic springs \cite{chandler1981exploiting}. Path-integral molecular dynamics (PIMD) \cite{ceperley_path_1995,tuckerman_efficient_1993} and path-integral Monte Carlo (PIMC) \cite{metropolis1953,hastings1970,herman_path_1982}  exploit this isomorphism to incorporate nuclear quantum effects (NQEs) rigorously, at $P$ times the cost of the corresponding classical simulation.

This cost has driven two broad strategies. The first retains the ring polymer but lowers the cost of each step or the number of steps needed to converge, through ring-polymer contraction \cite{markland_efficient_2008}, advanced integrators \cite{ceriotti2010efficient,liu_simple_2016}, or machine-learned force fields (MLFFs) \cite{fan2025performing,li2022using}. The second discards the ring polymer altogether, replacing it with a single classical system governed by an effective potential \cite{musil2022quantum,zaporozhets2024accurate} or driven by a colored-noise thermostat \cite{ceriotti2009nuclear}, generally at the price of approximations. Recently, GG-PI \cite{wang2026quantum} opened a third, generative route. It recognizes that the distribution of a single bead conditioned on its neighbors is the posterior of a Gaussian denoising problem, in which the noise variance reflects the quantum fluctuations. The denoiser can be trained on classical statistics alone and transfer across temperatures by adjusting the number of beads $P$ at a fixed imaginary-time slice $\tau=\beta/P$. In these methods, however, the remaining quantum context stays hard-wired. The nuclear masses and the coupling to a dissipative bath are fixed either in the simulation or, in GG-PI, in the very noise the model is trained to remove. Changing any one of them requires a new simulation or a retrained model, even though each enters the path measure only through the quadratic part of the action, which is known in closed form. Existing methods entangle this analytically known structure with the anharmonic classical statistics that must be learned.

Here we show that this structure can be exploited in full, so that nuclear quantum effects are injected entirely at sampling time rather than encoded in the learned model. The discretized imaginary-time action naturally separates into two parts: a quadratic term in the bead coordinates that encapsulates the complete quantum context in closed form, with the harmonic springs of the bare ring polymer as its simplest instance, and a residual potential term that factorizes into purely classical single-bead Boltzmann factors, blind to the quantum context. The entire quantum context enters the path measure as correlated Gaussian noise acting on otherwise independent classical replicas. Therefore, a denoiser trained to remove Gaussian noise from this classical distribution can be composed at sampling time with a matching quadratic component to recover the target path distribution exactly. This composition exists as long as the training noise does not exceed the intrinsic quantum uncertainty prescribed by the target ensemble, establishing a physical bound for the denoiser. Changing any element of the quantum context, including nuclear mass, bath coupling, and even the boundary conditions of the imaginary-time path, thus amounts to recomputing this quadratic component, with the denoiser left untouched. Mathematically, the composition is a real Hubbard–-Stratonovich \cite{stratonovich1957method,hubbard1959calculation} transformation applied not to the confining quadratic action, which admits no real decoupling field, but to its complement within the training noise, with the physical bound above as the positivity condition of that complement (\hyperref[app:hs]{End Matter}). The construction thereby offers an alternative to the imaginary-field route for confining quadratic forms, at the price of a bounded noise and a learnable residual.

We demonstrate the exact transferability of \method{} across temperature, isotope, bath coupling, and boundary conditions without retraining on three systems of increasing complexity. In a double-well coupled to a Caldeira--Leggett bath \cite{CaldeiraLeggett1983, Matsuo2008}, a single denoiser tracks numerically exact reference values as the temperature and dissipation strength are varied at sampling time. In the Zundel cation and in liquid water, the same construction reproduces proton delocalization and radial distribution functions across different temperatures and isotopic substitutions, in agreement with reference path-integral simulations. In water, opening the imaginary-time path of a tagged nucleus with the same denoiser further yields its end-to-end displacement and momentum distributions, which probe the thermal density matrix beyond the diagonal sampled by closed paths. The quantum context thereby becomes a property set at sampling time rather than one learned by the model: within each system, a single denoiser trained once supplies the classical Boltzmann statistics, while temperature, mass, dissipation, and the boundary conditions of the path enter through an analytic Gaussian component drawn at each sampling step. The same invariance extends beyond distinguishable particles, since bosonic exchange merely permutes the boundary conditions of the paths, entering through the analytic step (\hyperref[app:graph]{End Matter}).

\section{Method}

We consider a system of distinguishable particles in the canonical ($NVT$) ensemble at inverse temperature $\beta$. In the main text we present the primitive discretization together with an isotropic Gaussian noise model. 

Discretizing the imaginary time into $P$ slices of width $\tau=\beta/P$ maps the quantum Boltzmann distribution onto a ring polymer of $P$ replicas $\mathbf{x}=(\mathbf{x}_1,\dots,\mathbf{x}_P)$, $\mathbf{x}_k\in\mathbb{R}^{dN}$ with $\mathbf{x}_{P+1}\equiv\mathbf{x}_1$ \cite{chandler1981exploiting,herman_path_1982}. The discretized action splits into a quadratic and a residual part,
\begin{equation}
\begin{gathered}
  \pi(\mathbf{x})\;\propto\;
  \exp\!\Big[-\tfrac12\,\mathbf{x}^{\top}K\,\mathbf{x}\Big]\,
  \exp\!\big[-U(\mathbf{x})\big],\\
  U(\mathbf{x})=\tau\sum_{k=1}^{P}V(\mathbf{x}_k).
\end{gathered}
  \label{eq:split}
\end{equation}
The positive-semidefinite matrix $K$ collects the analytically known quadratic context. Its simplest instance is the bare ring polymer, where $\tfrac12\mathbf{x}^{\top}K\mathbf{x}=\frac{1}{2\hbar^{2}\tau}\sum_{k=1}^{P}(\mathbf{x}_{k+1}-\mathbf{x}_k)^{\top}M(\mathbf{x}_{k+1}-\mathbf{x}_k)$ is the harmonic spring energy coupling neighboring beads, with $M$ the diagonal mass matrix. For a system with additional harmonic terms, or one linearly coupled to a Caldeira--Leggett bath, $K$ is a more general positive-semidefinite matrix, still known in closed form. The residual $U$ collects the remaining single-bead potentials $V$, so that $e^{-U}=\prod_{k}e^{-\tau V(\mathbf{x}_k)}$ is a product of independent classical Boltzmann factors, all governed by the same potential $V$ at inverse temperature $\tau$.

We introduce an auxiliary path $\mathbf{y}=(\mathbf{y}_1,\dots,\mathbf{y}_P)$ and construct the joint distribution
\begin{equation}
\begin{gathered}
  p(\mathbf{x},\mathbf{y})=\pi(\mathbf{x})\,p(\mathbf{y}\mid\mathbf{x}),\\
  p(\mathbf{y}\mid\mathbf{x})=
  \mathcal{N}\!\big(\mathbf{y};\,(I-\sigma^{2}K)\mathbf{x},\;
  \sigma^{2}(I-\sigma^{2}K)\big).
\end{gathered}
  \label{eq:joint}
\end{equation}
For any normalized conditional, integrating out $\mathbf{y}$ returns the target, so the $\mathbf{x}$-marginal of Eq.~\eqref{eq:joint} is exact by construction. The Gaussian channel in Eq.~\eqref{eq:joint} is tuned so that the reverse conditional $p(\mathbf{x}\mid\mathbf{y})$ is freed of the quadratic context $K$.

Completing the square in $\mathbf{x}$ (\hyperref[app:square]{detailed in End Matter}) gives the reverse conditional
\begin{equation}
\begin{aligned}
  p(\mathbf{x}\mid\mathbf{y})
  &\;\propto\;
  \exp\!\Big[-\frac{\lVert\mathbf{x}-\mathbf{y}\rVert^{2}}{2\sigma^{2}}
  -U(\mathbf{x})\Big]\\[2pt]
  &\;=\;
  \prod_{k=1}^{P}
  \exp\!\Big[-\frac{\lVert \mathbf{x}_k-\mathbf{y}_k\rVert^{2}}{2\sigma^{2}}-\tau V(\mathbf{x}_k)\Big].
\end{aligned}
\label{eq:posterior}
\end{equation}
The quadratic context $K$ has cancelled entirely. Eq.~\eqref{eq:posterior} is exactly the Bayesian posterior of recovering a classical configuration $\mathbf{x}_k\sim e^{-\tau V}$ from an observation $\mathbf{y}_k=\mathbf{x}_k+\sigma\boldsymbol{\xi}$ corrupted by isotropic Gaussian noise---the denoising posterior underlying score-based and flow-based generative models \cite{song2020score,lipman_flow_2023}. Because $U$ is a sum of single-bead terms, this posterior factorizes over beads and one denoiser acting on a single replica suffices.

For the Gaussian channel in Eq.~\eqref{eq:joint} to be a valid distribution, its covariance $\sigma^{2}(I-\sigma^{2}K)$ must stay positive definite, which caps the noise variance:
\begin{equation}
  \sigma^{2}<\lambda_{\max}(K)^{-1}.
  \label{eq:ceiling}
\end{equation}
For the bare ring polymer, it reduces to $\sigma^{2}<\hbar^{2}\tau/4m$, one quarter of the free-particle mean-square imaginary-time displacement between adjacent slices \cite{tuckerman_statistical_2023}. Equation~\eqref{eq:ceiling} is the physical bound on the denoiser: the injected noise cannot exceed the quantum uncertainty encoded in the quadratic action, which is set by the stiffest mode.

We sample the joint density in Eq.~\eqref{eq:joint} by Gibbs sampling \cite{gelfand2000gibbs}, which constructs a Markov chain to iteratively draw each variable from its conditional distribution. In our case, the conditional distribution is given by Eq.~\eqref{eq:joint} and Eq.~\eqref{eq:posterior}. In each sweep, we alternate between two steps: (i) an analytic Gaussian draw $\mathbf{y}\sim p(\mathbf{y}\mid\mathbf{x})$ from Eq.~\eqref{eq:joint}, which injects the quantum context through $K$; and (ii) a bead-wise denoising draw $\mathbf{x}\sim p(\mathbf{x}\mid\mathbf{y})$ from Eq.~\eqref{eq:posterior}, supplied by the learned denoiser. The quantum context and the classical statistics thus enter through two separate, alternating updates.

This separation makes the construction transferable in a strong sense. For any family of quadratic contexts sharing the residual $V$, the slice $\tau$, and the noise level, the reverse conditional Eq.~\eqref{eq:posterior} is the same for every member (\hyperref[app:graph]{End Matter}). A single denoiser therefore samples an entire family of quantum ensembles, with temperature entering through the bead number $P$ at fixed $\tau$, isotopic mass through $M$ in $K$, dissipation through the bath kernel in $K$, and the boundary conditions of the path through the connectivity that $K$ encodes.

In practice we realize Eq.~\eqref{eq:posterior} with a conditional continuous normalizing flow trained by flow matching \cite{lipman_flow_2023} (\hyperref[app:network]{End Matter}). Since only the conditional $p(\mathbf{x}\mid\mathbf{y})$ is needed, training pairs may come from standard MD at inverse temperature $\tau$, restrained MD, or existing PIMD trajectories, all of which yield the same conditional (\hyperref[app:graph]{End Matter}).

\section{Results}

We validate the framework on three systems of increasing complexity. Within each system a single denoiser is used for all test conditions, at a fixed noise model whose per-component levels satisfy the ceiling Eq.~\eqref{eq:ceiling} throughout. The quantum context enters only through the analytic Gaussian step. In the Caldeira--Leggett double well the denoiser is numerically exact, giving a clean test of transfer across temperature and bath coupling. The Zundel cation uses a learned denoiser to test transfer across temperature and isotopic mass. In liquid water, we validate transfer across isotopic mass and use the same denoiser to sample open imaginary-time paths, testing transfer across the boundary conditions of the path.

\subsection{Dissipative double well}

We first validate our framework in the dissipative double well, a particle in a symmetric quartic double well $V(x)=(x^{2}-1)^{2}$ linearly coupled to an Ohmic bath of harmonic oscillators. This is a canonical benchmark for open quantum systems and the dissipation-driven quantum-to-classical crossover~\cite{CaldeiraLeggett1983,Matsuo2008}. Integrating out the bath couples the beads nonlocally while keeping the quadratic form:
\begin{equation}
\begin{aligned}
  \tfrac12\,\mathbf{x}^{\top}K\,\mathbf{x}
  &=
  \frac{m}{2\hbar^{2}\tau}\sum_{k}(x_{k+1}-x_{k})^{2}\\
  &\quad
  +\alpha\sum_{k<l}\frac{(\pi/P)^{2}}{4\sin^{2}[\pi(k-l)/P]}\,(x_{k}-x_{l})^{2}.
\end{aligned}
  \label{eq:cl_kernel}
\end{equation}
The first term is the ring-polymer springs and the second is the long-ranged imaginary-time kernel of the Ohmic bath~\cite{Matsuo2008}, whose strength $\alpha$ sets the dissipation. Crucially, $K=K_{\mathrm{spr}}+\alpha\,K_{\mathrm{bath}}$ is linear in $\alpha$. $\alpha=0$ is the isolated well, and increasing $\alpha$ strengthens the bath coupling and suppresses quantum delocalization between the wells. Because the system is one-dimensional, the single-bead posterior Eq.~\eqref{eq:posterior} is available by direct numerical quadrature, so the denoiser here is numerically exact. A single noise level $\sigma$, below the ceiling Eq.~\eqref{eq:ceiling} across test parameters, is used throughout. Changing $\alpha$ or the temperature then alters only the analytic Gaussian step.

\begin{figure}[hbt]
  \centering
  \includegraphics[width=0.98\columnwidth]{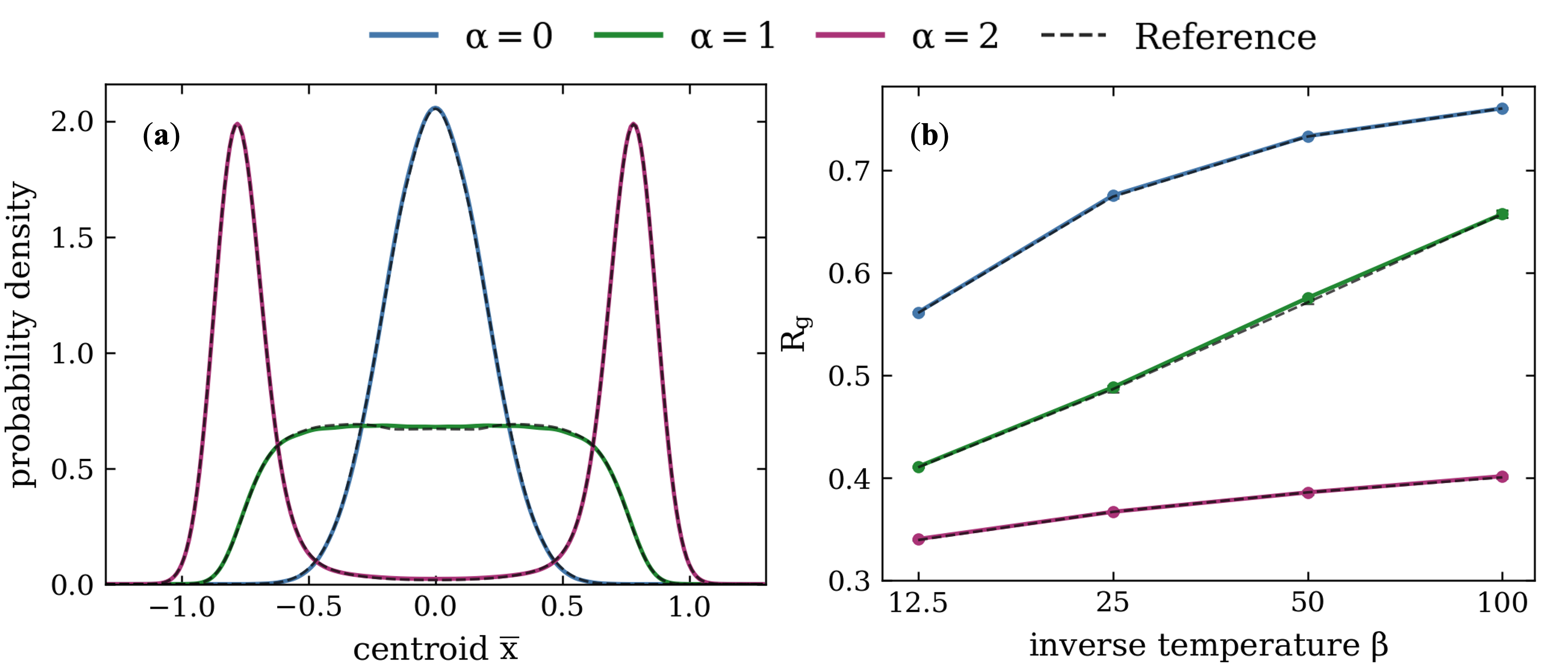}
  \caption{Bath and temperature transfer in the Caldeira--Leggett double well.
  (a) Centroid distribution for bath couplings $\alpha=0,1,2$. Increasing $\alpha$ drives the distribution from unimodal (the particle delocalized over both wells) to bimodal (localized in the wells).
  (b) Radius of gyration $R_g$ versus inverse temperature $\beta$ at fixed $\tau$ for the same couplings. $R_g$ increases as the temperature decreases and decreases with coupling.
  Solid: our method with a single denoiser; dashed: PIMC reference.}
  \label{fig:cl_main}
\end{figure}

Figure~\ref{fig:cl_main} shows that a single denoiser reproduces the PIMC reference across all three couplings and across temperature at fixed $\tau$.

\subsection{Zundel cation}

The Zundel cation $\mathrm{H_5O_2^+}$ describes a shared proton between two water molecules. It is a benchmark for nuclear quantum effects in hydrogen bonding and proton transfer~\cite{huang_ab_2005,suzuki_temperature_2013}. Here the denoiser is a learned model trained on existing PIMD trajectories at $300~\mathrm{K}$ with $P=32$, then applied across isotopes (H, D, T) and temperatures.

\begin{figure}[htb]
  \includegraphics[width=0.98\columnwidth]{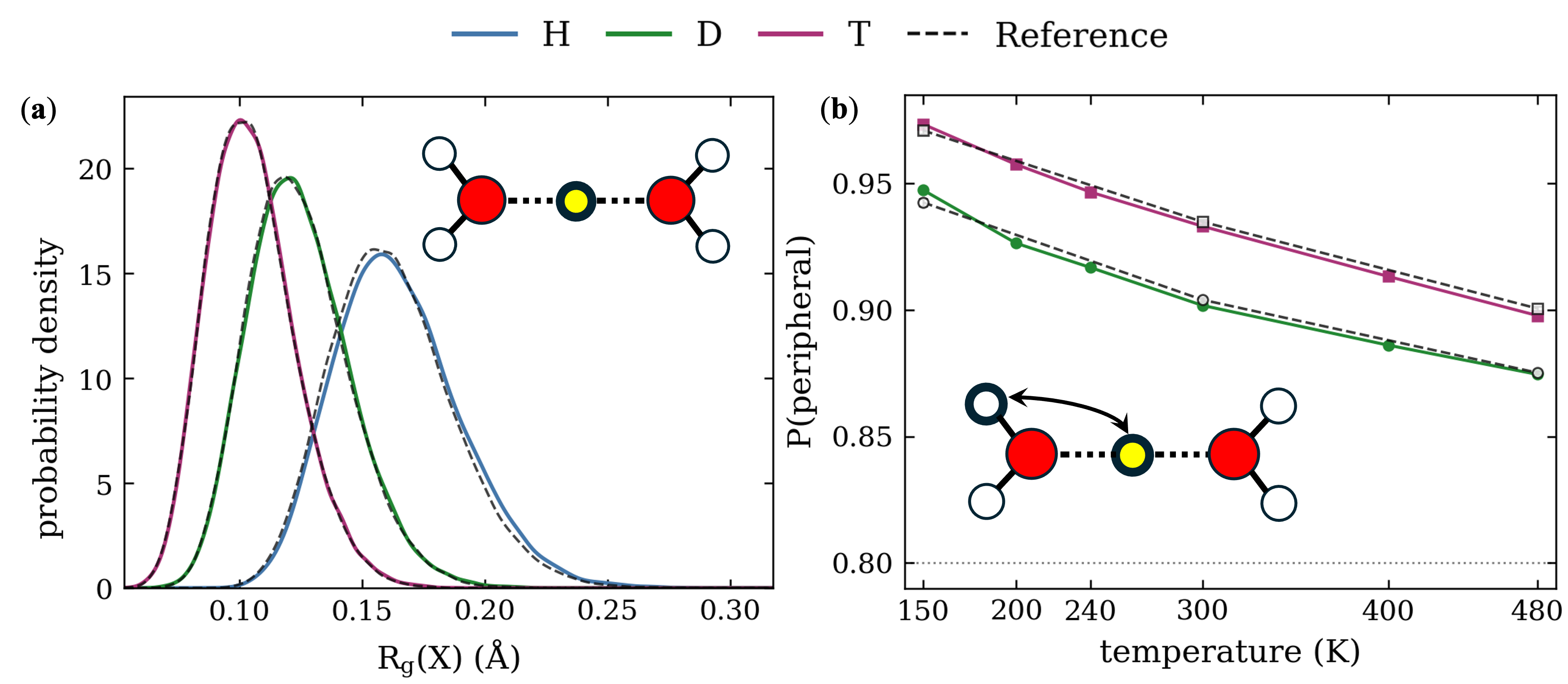}
  \caption{Isotope effects in the Zundel cation.
  (a) Radius of gyration $R_g$ of the shared nucleus for H, D, and T at $300~\mathrm{K}$ with $P=32$.
  (b) Probability that a substituted isotope occupies a peripheral site rather than the central shared site (highlighted in yellow in the illustration), versus temperature at fixed $\tau$. Heavier isotopes prefer the peripheral sites more strongly, and the preference weakens with temperature. The dotted line marks the isotope-free value $4/5$.
  Solid: our method with a single denoiser. Dashed: PIMD reference.}
  \label{fig:zundel_main}
\end{figure}

We examine two isotope effects. Fig.~\ref{fig:zundel_main}(a) reports the distribution of the radius of gyration $R_g$ of the shared nucleus for the three isotopes. Fig.~\ref{fig:zundel_main}(b) reports its site preference. On replacing one H by D or T, the heavier isotope may occupy one of the four peripheral $\mathrm{O\!-\!H}$ sites or the central shared site, and we compute its probability of being peripheral. Because mass enters only through $K$, this probability follows from an alchemical path in mass. We sample the corresponding sequence of $K$ matrices with the same denoiser and obtain the free energy between the shared and peripheral sites by MBAR \cite{shirts2008statistically}.

The shared-nucleus $R_g$ decreases from H to D to T (Fig.~\ref{fig:zundel_main}(a)), so heavier isotopes are less delocalized. Heavier isotopes also favor the peripheral sites over the shared site, and this preference grows with mass and weakens with temperature (Fig.~\ref{fig:zundel_main}(b)). Across isotopes and temperatures the single denoiser matches the PIMD reference.

\subsection{Liquid water}

Finally, we apply the framework to liquid water, $216$ molecules in a periodic box at $300~\mathrm{K}$ with $P=32$, described by the q-TIP4P/F force field~\cite{habershon_competing_2009}. Here the denoiser is trained from restrained MD. For each classical configuration $\mathbf{y}$, a short MD run restrained toward $\mathbf{y}$ draws a sample $\mathbf{x}$ from the posterior Eq.~\eqref{eq:posterior}, giving training pairs without any path-integral simulation. We use the single denoiser to examine isotope effects on the radial distribution functions (RDFs) of the O--O, O--H, and H--H pairs, on the intramolecular H--O--H angle, and, by opening the imaginary-time path of a tagged nucleus, on its end-to-end displacement and momentum distributions \cite{kapil2018anisotropy}.

\begin{figure}[htb]
  \includegraphics[width=0.98\columnwidth]{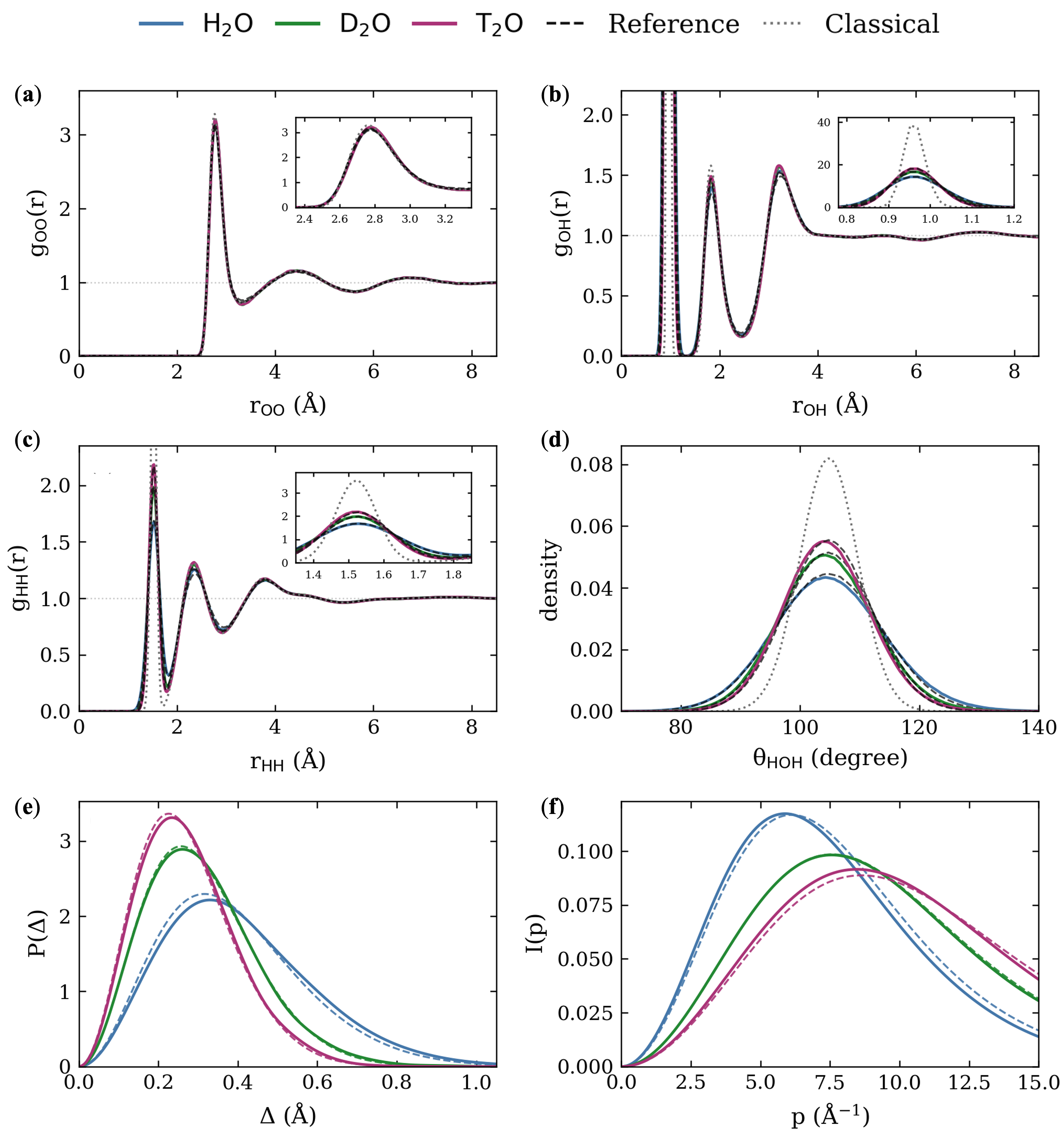}
  \caption{Isotope effects and open-path observables in liquid water.
  (a-c) Radial distribution functions of the O--O, O--H, and H--H pairs for H$_2$O, D$_2$O, and T$_2$O at $300~\mathrm{K}$ with $P=32$.
  (d) Intramolecular H--O--H angle distribution for the same isotopes.
  (e) Normalized radial end-to-end distribution $P(\Delta)$ of the opened imaginary-time path of a tagged nucleus for the same three liquids.
  (f) Radial momentum distribution $I(p)$ of the tagged nucleus.
  Solid: our method with a single denoiser. Dashed: PIMD reference, open-path PIMD in (e) and (f). Dotted: classical MD in (a-d).}
  \label{fig:water_main}
\end{figure}

Nuclear quantum effects broaden the distributions relative to classical MD, which is over-structured throughout (Fig.~\ref{fig:water_main}). The effect is strongest on the light hydrogens and weakens with isotope mass. The O--H covalent peak sharpens from H$_2$O to T$_2$O (Fig.~\ref{fig:water_main}(b), inset), the first H--H peak grows (Fig.~\ref{fig:water_main}(c), inset), and the H--O--H angle distribution narrows (Fig.~\ref{fig:water_main}(d)), while the O--O RDF is nearly isotope-independent (Fig.~\ref{fig:water_main}(a)). Across all isotopes the single denoiser matches the PIMD reference.

The open-path observables probe the quantum delocalization directly. The radial end-to-end distribution narrows from H$_2$O to T$_2$O (Fig.~\ref{fig:water_main}(e)), and the corresponding radial momentum distribution broadens and shifts to higher momentum (Fig.~\ref{fig:water_main}(f)). The same denoiser, recomposed with the open-chain quadratic context, follows the open-path PIMD reference for all three isotopes.

\section{Discussion}

\method{} partitions the path measure into an analytically known component and a learned component. The learned component is a single denoiser for the classical Boltzmann statistics, while temperature, isotopic mass, bath coupling, and the boundary conditions of the path enter through the analytic Gaussian step alone. Every demonstrated transfer is therefore the recomposition of a fixed denoiser with a different quadratic form $K$ under the ceiling Eq.~\eqref{eq:ceiling}. The underlying structure is the graph invariance of the reverse conditional (\hyperref[app:graph]{End Matter}). Once the residual and the noise model are fixed, every admissible quadratic context shares the same denoising posterior. The open-chain observables are the direct demonstration, the same denoiser yielding the end-to-end displacement and the momentum distribution once the edge closing the path of a tagged nucleus is deleted. A systematic wall-clock comparison is left to future work. Here we focus on the exactness and transferability of the construction.

Mathematically, our construction is a real Hubbard--Stratonovich (HS) transformation applied to the complement of the quantum coupling, $\sigma^{-2}I-K$, rather than to the confining coupling itself (\hyperref[app:hs]{End Matter}). Our construction and the HS transformation share the same localizing step. Conditioned on the auxiliary variable, the coupled system factorizes into subsystems, each linked only to its own component of the auxiliary. They differ in which coupling is transferred and in what becomes of the localized subsystems. In many-body applications the interaction is the intractable element. The transformation moves it into the field, the localized particles become conditionally Gaussian and are integrated out, and the difficulty reappears in the non-Gaussian marginal of the field. In the path measure the roles are reversed. The Gaussian part is precisely the quantum content and is known in closed form, while the intractable residual is classical and local in imaginary time. The channel accordingly transfers the known coupling. Each bead is then linked to its own auxiliary component through a conditional that is classical and non-Gaussian, the object the denoiser represents, and nothing is integrated out. The marginal of the auxiliary path is never required.

The same construction admits a second understanding at each sweep. The denoiser removes the full training noise every time it is invoked, while the channel injects only part of it, withholding a share corresponding to the quantum coupling (\hyperref[app:hs]{End Matter}, Eq.~\eqref{eq:em_budget}). The quantum context thus emerges from the ``over-denoising'' of the denoiser, and this leads to the ceiling in Eq.~\eqref{eq:ceiling}. This bound places GG-PI~\cite{wang2026quantum} outside the framework. GG-PI learns the conditional distribution of one bead given its two neighbors along the chain, which for the free ring polymer is Gaussian with variance $\hbar^{2}\tau/2m$, as the two adjacent springs contribute precision $2m/\hbar^{2}\tau$. This is twice the ceiling $\hbar^{2}\tau/4m$. The noise that GG-PI removes is thus the quantum fluctuation itself and exceeds every admissible training noise for the context it targets. No channel of the form Eq.~\eqref{eq:joint} exists at that level, no share can be withheld, and the quantum context cannot be separated from the learned conditional.

Our construction rests on elementary Gaussian identities, yet it becomes practical only once the denoising posterior Eq.~\eqref{eq:posterior} admits an amortized model. Score-based and flow-based generative models have recently made such models accurate and inexpensive to evaluate \cite{song2020score,lipman_flow_2023}. Since the Gaussian step is analytic, the accuracy of the sampled ensemble is determined entirely by the learned conditional. The framework thereby both demands high-fidelity generative models and provides a natural benchmark for their design, and a Metropolis correction can in principle remove the residual bias (\hyperref[app:mh]{End Matter}). More broadly, \method{} and our earlier work \cite{wang2026quantum, wang2026composing} follow one recipe, in which a model trained on easily sampled statistics is amortized and then extended to a much harder target by an augmented Gibbs sampler that supplies the analytically known structure at sampling time. The same learned model is reused across contexts, a philosophy that is portable to problems well beyond nuclear quantum effects.

As in any Markov chain method, mixing sets the remaining cost. Each sweep displaces the configuration on the scale of the training noise, which the ceiling ties to the stiffest mode, so the chain advances by local moves and collective rearrangements decorrelate slowly. Replica exchange across a ladder of admissible $K$ matrices at fixed bead number, all sampled with the same denoiser, provides a natural remedy. Because every member of the family shares the residual, the swap acceptance involves only the analytic quadratic forms, the residual canceling identically (\hyperref[app:repex]{End Matter}).

The latent-graph corollary (\hyperref[app:graph]{End Matter}) indicates the natural next step. Bosonic exchange becomes an analytic update over the permutation graphs that runs alongside the unchanged denoiser \cite{hirshberg2019path}, while fermionic signs remain outside the present framework \cite{li2019sign}. Beyond the path integral, the complement construction offers a general route for decoupling confining quadratic forms. The conventional imaginary-field transformation achieves this at the price of oscillatory weights, whereas the complement admits a real auxiliary variable at the price of the ceiling and of a residual that must be learnable.
\section*{Acknowledgments}
This work was supported by National Institutes of Health award R35 GM136381 and completed with computational resources administered by the University of Chicago Research Computing Center, including Beagle-3, a shared GPU cluster for biomolecular sciences supported by the NIH under the High-End Instrumentation (HEI) grant program award 1S10OD028655-0. JW's effort was supported by National Science Foundation award 2425899.

\bibliographystyle{apsrev4-2}
\bibliography{reference}

@book{horn2012matrix,
	title        = {Matrix analysis},
	author       = {Horn, Roger A and Johnson, Charles R},
	year         = 2012,
	publisher    = {Cambridge university press}
}

@article{fredrickson2002field,
  title={Field-theoretic computer simulation methods for polymers and complex fluids},
  author={Fredrickson, Glenn H and Ganesan, Venkat and Drolet, Fran{\c{c}}ois},
  journal={Macromolecules},
  volume={35},
  number={1},
  pages={16--39},
  year={2002},
  publisher={ACS Publications}
}

@article{wang2026composing,
  title={Composing diffusion priors with explicit physical context via generative Gibbs sampling},
  author={Wang, Weizhou and Weare, Jonathan and Dinner, Aaron R},
  journal={arXiv preprint arXiv:2605.10642},
  year={2026}
}

@article{li2022using,
	title        = {Using machine learning to greatly accelerate path integral ab initio molecular dynamics},
	author       = {Li, Chenghan and Voth, Gregory A},
	year         = 2022,
	journal      = {J. Chem. Theory Comput.},
	volume       = 18,
	number       = 2,
	pages        = {599--604}
}

@article{zaporozhets2024accurate,
	title        = {Accurate nuclear quantum statistics on machine-learned classical effective potentials},
	author       = {Zaporozhets, Iryna and Musil, F{\'e}lix and Kapil, Venkat and Clementi, Cecilia},
	year         = 2024,
	journal      = {J. Chem. Phys.},
	volume       = 161,
	number       = 13,
	pages        = 134102
}

@article{musil2022quantum,
	title        = {Quantum dynamics using path integral coarse-graining},
	author       = {Musil, F{\'e}lix and Zaporozhets, Iryna and No{\'e}, Frank and Clementi, Cecilia and Kapil, Venkat},
	year         = 2022,
	journal      = {J. Chem. Phys.},
	volume       = 157,
	number       = 18,
	pages        = 181102
}

@article{fan2025performing,
	title        = {Performing Path Integral Molecular Dynamics Using an Artificial Intelligence-Enhanced Molecular Simulation Framework},
	author       = {Fan, Cheng and Li, Maodong and Yuan, Sihao and Xie, Zhaoxin and Chen, Dechin and Yang, Yi Isaac and Gao, Yi Qin},
	year         = 2025,
	journal      = {J. Chem. Theory Comput.},
	volume       = 21,
	number       = 15,
	pages        = {7279--7289}
}

@article{hirshberg2019path,
	title        = {Path integral molecular dynamics for bosons},
	author       = {Hirshberg, Barak and Rizzi, Valerio and Parrinello, Michele},
	year         = 2019,
	journal      = {Proc. Natl. Acad. Sci. U.S.A.},
	volume       = 116,
	number       = 43,
	pages        = {21445--21449}
}

@article{chen2018neural,
	title        = {Neural ordinary differential equations},
	author       = {Chen, Ricky TQ and Rubanova, Yulia and Bettencourt, Jesse and Duvenaud, David K},
	year         = 2018,
	journal      = {Adv. Neural Inf. Process. Syst.},
	volume       = 31
}

@article{ceriotti2009nuclear,
	title        = {Nuclear quantum effects in solids using a colored-noise thermostat},
	author       = {Ceriotti, Michele and Bussi, Giovanni and Parrinello, Michele},
	year         = 2009,
	journal      = {Phys. Rev. Lett.},
	volume       = 103,
	number       = 3,
	pages        = {030603}
}

@article{wang2026quantum,
	title        = {Quantum statistics from classical simulations via generative Gibbs sampling},
	author       = {Wang, Weizhou and Zhang, Xuanxi and Weare, Jonathan and Dinner, Aaron R},
	year         = 2026,
	journal      = {arXiv preprint arXiv:2601.20228}
}

@article{li2019sign,
	title        = {Sign-problem-free fermionic quantum Monte Carlo: Developments and applications},
	author       = {Li, Zi-Xiang and Yao, Hong},
	year         = 2019,
	journal      = {Annual Review of Condensed Matter Physics},
	publisher    = {Annual Reviews},
	volume       = 10,
	number       = 1,
	pages        = {337--356}
}

@article{chandler1981exploiting,
	title        = {Exploiting the isomorphism between quantum theory and classical statistical mechanics of polyatomic fluids},
	author       = {Chandler, David and Wolynes, Peter G},
	year         = 1981,
	journal      = {J. Chem. Phys.},
	volume       = 74,
	number       = 7,
	pages        = {4078--4095}
}

@article{metropolis1953,
	title        = {Equation of State Calculations by Fast Computing Machines},
	author       = {Metropolis, N. and Rosenbluth, A. W. and Rosenbluth, M. N. and Teller, A. H. and Teller, E.},
	year         = 1953,
	journal      = {J. Chem. Phys.},
	volume       = 21,
	number       = 6,
	pages        = {1087--1092}
}

@article{hastings1970,
	title        = {Monte Carlo Sampling Methods Using Markov Chains and Their Applications},
	author       = {Hastings, W. K.},
	year         = 1970,
	journal      = {Biometrika},
	volume       = 57,
	number       = 1,
	pages        = {97--109}
}

@article{gelfand2000gibbs,
	title        = {Gibbs sampling},
	author       = {Gelfand, Alan E},
	year         = 2000,
	journal      = {J. Am. Stat. Assoc.},
	volume       = 95,
	number       = 452,
	pages        = {1300--1304}
}

@article{huang_ab_2005,
	title        = {\textit{{Ab} initio} potential energy and dipole moment surfaces for {H5O2}+},
	author       = {Huang, Xinchuan and Braams, Bastiaan J. and Bowman, Joel M.},
	year         = 2005,
	journal      = {J. Chem. Phys.},
	volume       = 122,
	number       = 4
}

@article{suzuki_temperature_2013,
	title        = {Temperature dependence on the structure of {Zundel} cation and its isotopomers},
	author       = {Suzuki, Kimichi and Tachikawa, Masanori and Shiga, Motoyuki},
	year         = 2013,
	journal      = {J. Chem. Phys.},
	volume       = 138,
	number       = 18
}

@article{habershon_competing_2009,
	title        = {Competing quantum effects in the dynamics of a flexible water model},
	author       = {Habershon, Scott and Markland, Thomas E. and Manolopoulos, David E.},
	year         = 2009,
	journal      = {J. Chem. Phys.},
	volume       = 131,
	number       = 2,
	pages        = {024501}
}

@article{ceriotti2010efficient,
	title        = {Efficient stochastic thermostatting of path integral molecular dynamics},
	author       = {Ceriotti, Michele and Parrinello, Michele and Markland, Thomas E and Manolopoulos, David E},
	year         = 2010,
	journal      = {J. Chem. Phys.},
	volume       = 133,
	number       = 12,
	pages        = 124104
}

@inproceedings{lipman_flow_2023,
	title        = {Flow Matching for Generative Modeling},
	author       = {Yaron Lipman and Ricky T. Q. Chen and Heli Ben-Hamu and Maximilian Nickel and Matthew Le},
	year         = 2023,
	booktitle    = {Int. Conf. Learn. Represent.}
}

@inproceedings{song2020score,
	title        = {Score-Based Generative Modeling through Stochastic Differential Equations},
	author       = {Yang Song and Jascha Sohl-Dickstein and Diederik P Kingma and Abhishek Kumar and Stefano Ermon and Ben Poole},
	year         = 2021,
	booktitle    = {Int. Conf. Learn. Represent.}
}

@article{liu_simple_2016,
	title        = {A simple and accurate algorithm for path integral molecular dynamics with the {Langevin} thermostat},
	author       = {Liu, Jian and Li, Dezhang and Liu, Xinzijian},
	year         = 2016,
	journal      = {J. Chem. Phys.},
	volume       = 145,
	number       = 2,
	pages        = {024103}
}

@article{ceperley_path_1995,
	title        = {Path integrals in the theory of condensed helium},
	author       = {Ceperley, D. M.},
	year         = 1995,
	journal      = {Rev. Mod. Phys.},
	volume       = 67,
	number       = 2,
	pages        = {279--355}
}

@book{tuckerman_statistical_2023,
	title        = {Statistical Mechanics: Theory and Molecular Simulation},
	shorttitle   = {Statistical mechanics},
	author       = {Tuckerman, Mark E.},
	year         = 2023,
	publisher    = {Oxford university press}
}

@article{herman_path_1982,
	title        = {On path integral {Monte} {Carlo} simulations},
	author       = {Herman, M. F. and Bruskin, E. J. and Berne, B. J.},
	year         = 1982,
	journal      = {J. Chem. Phys.},
	volume       = 76,
	number       = 10,
	pages        = {5150--5155}
}

@article{markland_nuclear_2018,
	title        = {Nuclear quantum effects enter the mainstream},
	author       = {Markland, Thomas E. and Ceriotti, Michele},
	year         = 2018,
	journal      = {Nat. Rev. Chem.},
	volume       = 2,
	number       = 3,
	pages        = {0109}
}

@article{tuckerman_efficient_1993,
	title        = {Efficient molecular dynamics and hybrid {Monte} {Carlo} algorithms for path integrals},
	author       = {Tuckerman, Mark E. and Berne, Bruce J. and Martyna, Glenn J. and Klein, Michael L.},
	year         = 1993,
	journal      = {J. Chem. Phys.},
	volume       = 99,
	number       = 4,
	pages        = {2796--2808}
}

@article{markland_efficient_2008,
	title        = {An efficient ring polymer contraction scheme for imaginary time path integral simulations},
	author       = {Markland, Thomas E. and Manolopoulos, David E.},
	year         = 2008,
	journal      = {J. Chem. Phys.},
	volume       = 129,
	number       = 2,
	pages        = {024105}
}

@article{shirts2008statistically,
	title        = {Statistically optimal analysis of samples from multiple equilibrium states},
	author       = {Shirts, Michael R and Chodera, John D},
	year         = 2008,
	journal      = {The Journal of chemical physics},
	publisher    = {AIP Publishing},
	volume       = 129,
	number       = 12
}

@inproceedings{stratonovich1957method,
	title        = {A method for the. computation of quantum distribution functions},
	author       = {Stratonovich, Ruslan Leont'evich},
	year         = 1957,
	booktitle    = {Doklady Akademii Nauk},
	volume       = 115,
	number       = 6,
	pages        = {1097--1100},
	organization = {Russian Academy of Sciences}
}

@article{hubbard1959calculation,
	title        = {Calculation of partition functions},
	author       = {Hubbard, John},
	year         = 1959,
	journal      = {Physical Review Letters},
	publisher    = {APS},
	volume       = 3,
	number       = 2,
	pages        = 77
}

@article{kapil2018anisotropy,
	title        = {Anisotropy of the proton momentum distribution in water},
	author       = {Kapil, Venkat and Cuzzocrea, Alice and Ceriotti, Michele},
	year         = 2018,
	journal      = {The Journal of Physical Chemistry B},
	publisher    = {ACS Publications},
	volume       = 122,
	number       = 22,
	pages        = {6048--6054}
}

@article{CaldeiraLeggett1983,
	title        = {Quantum tunnelling in a dissipative system},
	author       = {Caldeira, A. O. and Leggett, A. J.},
	year         = 1983,
	journal      = {Ann. Phys.},
	volume       = 149,
	number       = 2,
	pages        = {374--456},
	doi          = {10.1016/0003-4916(83)90202-6}
}

@article{Matsuo2008,
	title        = {Quantum-classical transition and decoherence in dissipative double-well potential systems: {Monte} {Carlo} algorithm},
	author       = {Matsuo, Takeshi and Natsume, Yuhei and Kato, Takeo},
	year         = 2008,
	journal      = {Phys. Rev. B},
	volume       = 77,
	pages        = 184304,
	doi          = {10.1103/PhysRevB.77.184304}
}

\clearpage
\appendix


\section{Derivation of the reverse conditional}
\label{app:square}
Let $A\equiv I-\sigma^{2}K$, so that the channel in Eq.~\eqref{eq:joint} reads $p(\mathbf{y}\mid\mathbf{x})=\mathcal{N}(\mathbf{y};\,A\mathbf{x},\,\sigma^{2}A)$. Under the ceiling Eq.~\eqref{eq:ceiling}, $A$ is symmetric positive definite, so $A^{-1}$ exists. Since $p(\mathbf{x}\mid\mathbf{y})\propto\pi(\mathbf{x})\,p(\mathbf{y}\mid\mathbf{x})$ at fixed $\mathbf{y}$, we track only the $\mathbf{x}$-dependence, and $c(\mathbf{y})$ collects $\mathbf{x}$-independent terms and may change between lines. Using the symmetry of $A$, we expand the exponent of the Gaussian
channel as
\begin{equation}
  (\mathbf{y}-A\mathbf{x})^{\top}A^{-1}(\mathbf{y}-A\mathbf{x})
  =\mathbf{x}^{\top}\!A\,\mathbf{x}-2\,\mathbf{y}^{\top}\mathbf{x}
  +\mathbf{y}^{\top}\!A^{-1}\mathbf{y}.
  \label{eq:em_expand}
\end{equation}
Substituting this expansion into the logarithm of $p(\mathbf{x}\mid\mathbf{y})$ gives
\begin{equation}
  \begin{aligned}
    \log p(\mathbf{x}\mid\mathbf{y})
    &=-\frac{\mathbf{x}^{\top}\!A\mathbf{x}-2\,\mathbf{y}^{\top}\mathbf{x}}{2\sigma^{2}}
      -\frac{\mathbf{x}^{\top}K\mathbf{x}}{2}-U(\mathbf{x})+c(\mathbf{y})\\
    &=-\frac{\lVert\mathbf{x}\rVert^{2}-2\,\mathbf{y}^{\top}\mathbf{x}}{2\sigma^{2}}
      -U(\mathbf{x})+c(\mathbf{y})\\
    &=-\frac{\lVert\mathbf{x}-\mathbf{y}\rVert^{2}}{2\sigma^{2}}
      -U(\mathbf{x})+c(\mathbf{y}),
  \end{aligned}
\label{eq:em_square}
\end{equation}
using $A+\sigma^{2}K=I$ in the second line.

\section{Complement Hubbard--Stratonovich transformation}
\label{app:hs}

For a symmetric positive-definite matrix $B$, the Hubbard--Stratonovich (HS) identity reads
\begin{equation}
  e^{\frac12\mathbf{x}^{\top}B\,\mathbf{x}}
  =\mathbb{E}_{\mathbf{z}\sim\mathcal{N}(0,B)}\!\big[e^{\mathbf{x}^{\top}\mathbf{z}}\big],
  \label{eq:em_hs}
\end{equation}
a Gaussian mixture of linear tilts whose covariance enters the exponent with a positive sign. However, the quadratic action of the path integral is confining, $e^{-\frac12\mathbf{x}^{\top}K\mathbf{x}}$ with $K\succeq0$, and decoupling it through Eq.~\eqref{eq:em_hs} requires an imaginary coupling $i\,\mathbf{x}^{\top}\mathbf{z}$. This produces oscillatory weights and underlies the sign problem of auxiliary-field methods \cite{fredrickson2002field}.

The channel Eq.~\eqref{eq:joint} circumvents this obstruction by transforming the complement of the coupling within the training noise. Writing
\begin{equation}
  \begin{aligned}
  e^{-\frac12\mathbf{x}^{\top}K\mathbf{x}}
  &=e^{-\frac{\lVert\mathbf{x}\rVert^{2}}{2\sigma^{2}}}\,
  e^{+\frac12\mathbf{x}^{\top}B\,\mathbf{x}},\\
  B&=\sigma^{-2}I-K,
  \end{aligned}
\label{eq:em_complement}
\end{equation}
the complement $B$ is positive definite exactly under the ceiling Eq.~\eqref{eq:ceiling}. Applying Eq.~\eqref{eq:em_hs} to $B$ leads to the joint distribution
\begin{equation}
  p(\mathbf{x},\mathbf{z})\propto
  \exp\!\Big[-\frac{\lVert\mathbf{x}\rVert^{2}}{2\sigma^{2}}
  +\mathbf{x}^{\top}\mathbf{z}-U(\mathbf{x})
  -\tfrac12\,\mathbf{z}^{\top}B^{-1}\mathbf{z}\Big],
  \label{eq:em_hsjoint}
\end{equation}
whose $\mathbf{z}$-marginal recovers $\pi(\mathbf{x})$ by Eq.~\eqref{eq:em_hs}. In the rescaled variable $\mathbf{y}=\sigma^{2}\mathbf{z}$ the joint becomes
\begin{equation}
  p(\mathbf{x},\mathbf{y})\propto
  \exp\!\Big[-\frac{\lVert\mathbf{x}\rVert^{2}
  -2\,\mathbf{x}^{\top}\mathbf{y}}{2\sigma^{2}}
  -U(\mathbf{x})
  -\tfrac12\,\mathbf{y}^{\top}\Sigma^{-1}\mathbf{y}\Big],
  \label{eq:em_yjoint}
\end{equation}
with $\Sigma=\sigma^{4}B=\sigma^{2}(I-\sigma^{2}K)$. Completing the square in $\mathbf{y}$ at fixed $\mathbf{x}$, and in $\mathbf{x}$ at fixed $\mathbf{y}$, shows that the two conditionals of Eq.~\eqref{eq:em_yjoint} are exactly the channel Eq.~\eqref{eq:joint} and the posterior Eq.~\eqref{eq:posterior}. The construction of the main text is therefore a real HS transformation applied to the complement of the quantum coupling.

This construction has an operational reading. In the absence of quadratic context, $K=0$, the channel injects the full training noise, $\mathbf{y}\sim\mathcal{N}(\mathbf{x},\sigma^{2}I)$, so each sweep corrupts the configuration with exactly the noise the denoiser was trained to remove, and the loop leaves the classical product distribution $e^{-U}$ invariant. A nonzero quantum context contracts the channel mean by $I-\sigma^{2}K$ and withholds the matching share of the training noise
\begin{equation}
  \underbrace{\sigma^{2}\big(I-\sigma^{2}K\big)}_{\text{injected}}
  \;+\;
  \underbrace{\sigma^{4}K}_{\text{withheld}}
  \;=\;
  \underbrace{\sigma^{2}I}_{\text{training noise}}.
  \label{eq:em_budget}
\end{equation}
The denoiser removes the full training noise in every sweep, while the channel injects only the remainder. This deficit, which encodes the chosen quantum coupling, turns the stationary distribution from the classical product $e^{-U}$ into the path measure $\pi$. Because the injected covariance cannot be negative, the withheld share can never exceed the training noise; this constraint is the ceiling Eq.~\eqref{eq:ceiling}.


\section{Graph-invariant denoising decomposition}
\label{app:graph}

The main-text construction is one instance of a decomposition defined on a family of graphs. Let $G=(V,E)$ be a graph whose vertices carry the configurations $\mathbf{x}=(\mathbf{x}_1,\dots,\mathbf{x}_B)$. Each vertex $\mathbf{x}_b$ is the unit on which the denoiser acts, and the edges encode analytically known quadratic couplings. The target distribution associated with $G$ is
\begin{equation}
  \pi_G(\mathbf{x})
  \propto\exp\!\Big[-\tfrac12\,\mathbf{x}^{\top}K_G\,\mathbf{x}\Big]\exp[-U(\mathbf{x})],
  \label{eq:em_target}
\end{equation}
where $K_G=K_G^{\top}\succeq0$ is the quadratic context that collects the edge couplings and any on-site terms,
\begin{equation}
  \tfrac12\,\mathbf{x}^{\top}K_G\,\mathbf{x}
  =\sum_{(b,b')\in E} \tfrac12\,k_{bb'}\lVert\mathbf{x}_b-\mathbf{x}_{b'}\rVert^{2}
  +\tfrac12\,\mathbf{x}^{\top}K^{(2)}\mathbf{x},
  \label{eq:em_KG}
\end{equation}
with $k_{bb'}$ the stiffness of edge $(b,b')$ and $K^{(2)}$ any remaining analytically known contribution, such as a harmonic restraint or a nonlocal bath kernel. Throughout, $G$ denotes the complete graph-structured quadratic context, comprising the connectivity, the edge weights, the on-site terms, and the number of vertices. A linear term in the action would require a corresponding shift of the channel mean, and we omit it for simplicity.

Fix a residual $U$ and a noise matrix $R=\mathrm{diag}(R_1,\dots,R_B)$, which is block-diagonal with one block $R_b\succ0$ per vertex, both shared by every member of a family $\mathcal{G}$ of such contexts. For each $G$ the auxiliary configuration $\mathbf{y}=(\mathbf{y}_1,\dots,\mathbf{y}_B)$ is drawn from the channel
\begin{equation}
  p(\mathbf{y}\mid\mathbf{x},G)=
  \mathcal{N}\!\big(\mathbf{y};\,(I-RK_G)\mathbf{x},\;R-RK_G R\big),
  \label{eq:em_channel}
\end{equation}
which reduces to Eq.~\eqref{eq:joint} for $R=\sigma^2 I$. The channel is a valid distribution when $R-RK_G R\succ0$, equivalently
\begin{equation}
  \lambda_{\max}\!\big(R^{1/2}K_G R^{1/2}\big)<1,
  \label{eq:em_ceiling}
\end{equation}
which recovers $\sigma^2<\lambda_{\max}(K_G)^{-1}$ in the isotropic case. Since the channel is normalized, integrating out $\mathbf{y}$ returns $\pi_G$ for every $G$, so the augmentation is exact by construction.

\medskip
\noindent\textit{Proposition (graph-invariant reverse conditional).}
Let the residual $U$ be independent of the graph $G$, let the noise matrix $R$ be common to every member of $\mathcal{G}$, and let the ceiling Eq.~\eqref{eq:em_ceiling} hold for every $G\in\mathcal{G}$. Then for every $G\in\mathcal{G}$, the conditional distribution of $\mathbf{x}$ given $\mathbf{y}$ under the joint $\pi_G(\mathbf{x})\,p(\mathbf{y}\mid\mathbf{x},G)$ is
\begin{equation}
  p(\mathbf{x}\mid\mathbf{y})\propto
  \exp\!\Big[-\tfrac12(\mathbf{x}-\mathbf{y})^{\top}R^{-1}(\mathbf{x}-\mathbf{y})-U(\mathbf{x})\Big],
  \label{eq:em_posterior}
\end{equation}
independent of $G$. A single denoiser trained for Eq.~\eqref{eq:em_posterior} therefore serves the entire family, whose members differ only through their analytic channels Eq.~\eqref{eq:em_channel}. The proof repeats the completion of the square in Eqs.~\eqref{eq:em_expand} and \eqref{eq:em_square} with $A=I-RK_G$; the channel mean and covariance share the factor $A$, so $K_G$ cancels.

When $U$ separates over the vertices, $U(\mathbf{x})=\sum_b U_b(\mathbf{x}_b)$, the block-diagonal structure of $R$ factorizes the posterior,
\begin{equation}
\begin{gathered}
p(\mathbf{x}\mid\mathbf{y})
 =\prod_{b=1}^{B}p_b(\mathbf{x}_b\mid\mathbf{y}_b),\\
p_b
 \propto
\exp\!\Big[-\tfrac12(\mathbf{x}_b-\mathbf{y}_b)^{\top}R_b^{-1}(\mathbf{x}_b-\mathbf{y}_b)-U_b(\mathbf{x}_b)\Big].
\end{gathered}
\label{eq:em_factor}
\end{equation}
Each $p_b$ is the denoising posterior of a single vertex drawn from $e^{-U_b}$ and corrupted with covariance $R_b$. Members of $\mathcal{G}$ may also differ in size. A change of temperature at fixed $\tau$ changes the number of vertices, and the invariant object is then the single-vertex conditional $p_b$, reused across vertex counts.

Training pairs may be generated from any convenient member of the family. At one extreme, $K_G=0$ reduces sampling to standard MD followed by Gaussian corruption; at the other, the analytic channel converts existing PIMD/PIMC paths at the same $\tau$ into vertex pairs. Alternatively, restrained MD samples $p_b(\mathbf{x}_b\mid\mathbf{y}_b)$ directly. Although the marginal of $\mathbf{y}_b$ does not change the target conditional, its coverage affects denoiser accuracy in practice.

The uniform ceiling is often inherited automatically. Deleting an edge $(b,b')$ subtracts the positive-semidefinite harmonic term $\tfrac12\,k_{bb'}\lVert\mathbf{x}_b-\mathbf{x}_{b'}\rVert^{2}$ from Eq.~\eqref{eq:em_KG}, which lowers the context matrix in the Loewner partial order, $K_{G'}\preceq K_G$. Because congruence preserves this order and the largest eigenvalue is monotonic on it~\cite{horn2012matrix}, it follows that $\lambda_{\max}\!\big(R^{1/2}K_{G'}R^{1/2}\big)\le \lambda_{\max}\!\big(R^{1/2}K_{G}R^{1/2}\big)$. In physical terms, removing a spring can only soften the system; in mathematical terms, every subgraph of an admissible graph remains strictly admissible at the same noise level.

\medskip
\noindent\textit{Corollary (latent graph).}
Let the graph itself be a random variable with nonnegative weights $p(G)$ and unnormalized joint density
\begin{equation}
p(\mathbf{x},G)\propto
p(G)\,\exp\!\Big[-\tfrac12\,\mathbf{x}^{\top}K_G\,\mathbf{x}-U(\mathbf{x})\Big],
\label{eq:em_latent}
\end{equation}
normalized over $\mathbf{x}$ and $G$ together. Let Eq.~\eqref{eq:em_ceiling} hold for every $G$ in the support of $p(G)$. Augmenting each graph with its channel
\begin{equation}
  p(\mathbf{x},\mathbf{y},G)=p(\mathbf{x},G)\,p(\mathbf{y}\mid\mathbf{x},G),
\end{equation}
and completing the square as above separates $\mathbf{x}$ from $G$,
\begin{equation}
\begin{aligned}
p(\mathbf{x},\mathbf{y},G)&\propto f(\mathbf{x},\mathbf{y})\,h(\mathbf{y},G),\\
f(\mathbf{x},\mathbf{y})&=\exp\!\Big[-\tfrac12(\mathbf{x}-\mathbf{y})^{\top}R^{-1}(\mathbf{x}-\mathbf{y})-U(\mathbf{x})\Big],\\
h(\mathbf{y},G)&=p(G)\,{\det}(I-K_GR)^{-1/2}\\
&\quad \cdot\,e^{-\frac12\,\mathbf{y}^{\top}(I-K_GR)^{-1}K_G\,\mathbf{y}}.
\end{aligned}
\label{eq:em_fh}
\end{equation}
Hence $\mathbf{x}\perp G\mid\mathbf{y}$, and $p(\mathbf{x},G\mid\mathbf{y})=p(\mathbf{x}\mid\mathbf{y})\,p(G\mid\mathbf{y})$ with $p(G\mid\mathbf{y})\propto h(\mathbf{y},G)$. Given the auxiliary configuration $\mathbf{y}$, the configuration update through the graph-blind denoiser and the graph update through $p(G\mid\mathbf{y})$ therefore proceed independently and in parallel. The auxiliary configuration screens the physical coordinates from the quadratic context entirely.

In the path integral of the main text, each vertex carries one replica $\mathbf{x}_b\equiv\mathbf{x}_k\in\mathbb{R}^{dN}$ with $U_b=\tau V(\mathbf{x}_k)$, and $G$ is the imaginary-time graph of the ring polymer, with a Gaussian bath entering as additional nonlocal edges. Within each transfer family, the noise matrix $R_b$ is fixed once and shared by all contexts; only the quadratic context $K_G$ changes. Opening the imaginary-time path of a tagged atom deletes the edge that closes its cycle. We adopt the factorization of Ref.~\cite{kapil2018anisotropy}, in which every bead retains the residual weight $\tau V$, so $U$ is common to the open and the closed graph, the proposition applies, and the ceiling is inherited by the monotonicity above.

Bosonic exchange assigns uniform weights $p(G)$ to the permutation graphs that reconnect the imaginary-time endpoints, and the corollary reduces its sampling to the analytic update
$p(G\mid\mathbf{y})\propto h(\mathbf{y},G)$ of Eq.~\eqref{eq:em_fh}
alongside the unchanged denoiser. In principle, the same denoiser combined with this permutation update therefore samples bosonic systems. Fermionic weights carry signs, outside the present hypotheses, and are left to future work.

\subsection{Conditional normalizing flow}
\label{app:network}

We realize each single-vertex posterior $p_b(\mathbf{x}_b\mid\mathbf{y}_b)$ of Eq.~\eqref{eq:em_factor} with a conditional continuous normalizing flow (CNF) trained by flow matching~\cite{lipman_flow_2023,chen2018neural}. Conditioned on $\mathbf{y}_b$, the model transports a Gaussian base sample $\mathbf{x}^{0}\sim\mathcal{N}(\mathbf{y}_b,R_b)$ to a target sample $\mathbf{x}^{1}\sim p_b(\mathbf{x}_b\mid\mathbf{y}_b)$ along $s\in[0,1]$ through the ODE
\begin{equation}
  \frac{d\mathbf{x}^{s}}{ds}=v_\theta(\mathbf{x}^{s},\mathbf{y}_b,s),
  \qquad \mathbf{x}^{s=0}=\mathbf{x}^{0}.
\label{eq:em_ode}
\end{equation}
The conditional is equivariant under simultaneous rigid transformations of the sample and conditioning configuration. We construct the velocity field from relative coordinates so that, for $Q\in O(3)$ and $\mathbf{t}\in\mathbb{R}^{3}$,
\begin{equation}
  v_\theta(Q\mathbf{x}+\mathbf{t},Q\mathbf{y}_b+\mathbf{t},s)
  =Q\,v_\theta(\mathbf{x},\mathbf{y}_b,s).
  \label{eq:em_equiv}
\end{equation}
Because the Gaussian base is centered at $\mathbf{y}_b$, translating $\mathbf{y}_b$ translates the initial condition and the entire ODE trajectory by the same $\mathbf{t}$.
A lightweight network suffices, as the base is already centered at $\mathbf{y}_b$ and localized by $R_b$. The velocity is trained by the conditional flow-matching loss
\begin{equation}
\begin{aligned}
\mathcal{L}(\theta)
&=\mathbb{E}_{s,\,(\mathbf{x}^{0},\mathbf{x}^{1})}
\big\lVert
v_\theta\big((1-s)\mathbf{x}^{0}+s\mathbf{x}^{1},\,\mathbf{y}_b,\,s\big)\\
&\qquad-(\mathbf{x}^{1}-\mathbf{x}^{0})
\big\rVert^{2},
\end{aligned}
\label{eq:em_fmloss}
\end{equation}
with $s\sim\mathcal{U}[0,1]$.

\subsection{Flow likelihood and Metropolis-Hastings correction}
\label{app:mh}

The conditional flow has an exact likelihood. Integrating the instantaneous change of variables along the ODE gives
\begin{equation}
\log q_\theta(\mathbf{x}_b\mid\mathbf{y}_b)
=\log\mathcal{N}(\mathbf{x}^{0};\,\mathbf{y}_b,R_b)
-\int_{0}^{1}\nabla\!\cdot v_\theta(\mathbf{x}^{s},\mathbf{y}_b,s)\,ds,
\label{eq:likelihood}
\end{equation}
where $\mathbf{x}^{0}$ is obtained by integrating the ODE backward from $\mathbf{x}_b$. An independence Metropolis--Hastings step then removes the residual bias of the learned conditional. A proposal  $\mathbf{x}_b'\sim q_\theta(\cdot\mid\mathbf{y}_b)$ is accepted with probability
\begin{equation}
\alpha=\min\!\left\{1,\;
\frac{\tilde p_b(\mathbf{x}_b'\mid\mathbf{y}_b)\,
q_\theta(\mathbf{x}_b\mid\mathbf{y}_b)}
{\tilde p_b(\mathbf{x}_b\mid\mathbf{y}_b)\,
q_\theta(\mathbf{x}_b'\mid\mathbf{y}_b)}\right\},
\label{eq:mhacc}
\end{equation}
with $\tilde p_b$ the unnormalized posterior. This can remove the error in the learned conditional at the price of extra potential evaluations per proposal. The reported results do not employ the correction.

\subsection{Replica exchange}
\label{app:repex}
By the \hyperref[app:graph]{graph-invariant proposition}, many ensembles can share the same denoiser with different $K$. Suppose two ensembles $\pi_i$ and $\pi_j$ have the same residual $U$ but different quadratic contexts $K_i$ and $K_j$. Exchanging their configurations is accepted with probability $\min(1,e^{\Delta_{ij}})$, where
\begin{equation}
\Delta_{ij}=\tfrac12\big(
\mathbf{x}_i^{\top}K_i\mathbf{x}_i
+\mathbf{x}_j^{\top}K_j\mathbf{x}_j
-\mathbf{x}_i^{\top}K_j\mathbf{x}_i
-\mathbf{x}_j^{\top}K_i\mathbf{x}_j\big),
\label{eq:swap}
\end{equation}
in which $U$ has cancelled identically. The acceptance involves only the analytic quadratic forms. This provides a way to exchange across mass ladders to accelerate the mixing. Exchange across temperature changes the bead number and is not of this form.


\end{document}